\documentclass{PoS}
\usepackage{amsmath}
\usepackage{amssymb,amsthm}
\usepackage{slashed}
\usepackage{graphicx}

\newcommand{\PZ}{\text{Z}}

\newcommand{\PW}{\text{W}}

\newcommand{\rT}{\mathrm{T}}

\newcommand{\rd}{\mathrm{d}}

\title{Vector-boson pair production at the LHC:  \\  
                              Electroweak corrections in HERWIG++}
\ShortTitle{Electroweak corrections to vector-boson pair production in HERWIG++}

\author{Stefan Gieseke\\
        Karlsruhe Institute of Technology (KIT), 
        Institut f\"ur Theoretische Physik,\\
        D-76128 Karlsruhe, Germany \\ 
        E-mail: \email{Stefan.Gieseke@kit.edu}}
\author{\speaker{Tobias Kasprzik}\\
        Karlsruhe Institute of Technology (KIT), 
        Institut f\"ur Theoretische Teilchenphysik, \\        
        D-76128 Karlsruhe, Germany\\
        E-mail: \email{Tobias.Kasprzik@kit.edu}}
\author{Johann H. K\"uhn \\ 
        Karlsruhe Institute of Technology (KIT), 
        Institut f\"ur Theoretische Teilchenphysik,\\
        D-76128 Karlsruhe, Germany \\ 
        E-mail: \email{Johann.Kuehn@kit.edu}}

      \abstract{Vector-boson pair production is of great
        phenomenological importance at the LHC. These processes will
        help to validate the Standard Model at highest energies, and
        they may also open the door for the discovery of new physics
        potentially showing up in subtle modifications of the
        non-abelian structure of weak interactions. In this letter, we
        review the status of the corresponding theory predictions,
        focusing on the higher-order electroweak corrections. We
        present a NLO analysis of electroweak corrections to W-pair,
        W$^+$Z and Z-pair production at the LHC, including all mass
        effects as well as leptonic decays. Contributions of
        photon-induced processes and massive-boson radiation are also
        discussed. The electroweak corrections are implemented in the
        HERWIG++ Monte Carlo generator, where they are combined with
        QCD corrections. We also propose a simple and straight-forward
        method allowing for an \emph{a posteriori} implementation of
        electroweak corrections to vector-boson pair production in any
        Monte Carlo event sample.

\begin{flushright}
\emph{TTP13-030 \\ 
KA-TP-30-2013\\
LPN13-070 \\ 
SFB/CPP-13-72}
\end{flushright}}

\FullConference{EPS HEP,\\
		18-24 July 2013,\\
		Stockholm, Sweden}

\begin{document}
\section{Introduction}
As stated in the abstract, a profound understanding of vector-boson pair
production processes at the LHC is desirable for various reasons.
Consequently, great effort has been made during the last years to push
the theory predictions for this process class to a new level, where,
besides the dominating next-to-leading-order (NLO) QCD corrections (see,
e.g., Ref.~\cite{Bierweiler:2012kw} and references therein), also
electroweak (EW) effects have been studied extensively. In particular,
the interplay of EW corrections and anomalous couplings has been
investigated in Ref.~\cite{Accomando:2005xp}
in the high-energy limit, including leptonic decays and off-shell
effects. Very recently, EW corrections to W-pair production have been
computed in the double-pole approximation (DPA), even taking into
account the full mass dependence~\cite{Billoni:2013aba}. Leading
two-loop effects at high transverse momenta were discussed in
Ref.~\cite{Kuhn:2011mh} for W-pairs. A detailed analysis of on-shell
$V$-boson pair production ($V=\PW^\pm,\PZ$) including EW corrections has
been provided in Refs.~\cite{Bierweiler:2012kw, Bierweiler:2013dja},
consistently including all mass effects. A detailed review of NLO
effects in pair production of massive bosons can also be found in
Ref.~\cite{Baglio:2013toa}, emphasizing the importance of photon-induced
contributions. A brief discussion of the phenomenological implications
of those effects will be given in section~\ref{se:onshell}.

In addition to NLO corrections, first steps towards the computation of
NNLO QCD corrections to massive $V$-pair production have been
taken~\cite{Gehrmann:2013cxs}, and approximate NNLO results for
$\mathrm{W^+Z}$ and $\mathrm{WW}$ production have been provided for
high-$p_\rT$ observables~\cite{Campanario:2012fk}, as well as for WW
production in the threshold limit~\cite{Dawson:2013lya}.

Nevertheless, a phenomenological analysis of EW corrections in $V$-pair
production at the LHC, including leptonic decays \emph{and} mass effects,
is still missing for the ZZ and WZ channels. This gap is partly closed
in~section~\ref{se:off-shell}, where first results of resonant four-lepton
production at NLO EW accuracy are presented. To facilitate the
phenomenological analysis of data including EW corrections, a straight
forward implementation of EW corrections in the
HERWIG++~\cite{Bahr:2008pv} setup is proposed in
section~\ref{se:herwig}, combining the flexibility of a state-of-the-art
Monte Carlo (MC) generator with EW precision.

\section{Electroweak Corrections at NLO}
\subsection{On-shell gauge-boson pair production}
\label{se:onshell}
\begin{figure}
\begin{center}
\includegraphics[width = 0.9\textwidth]{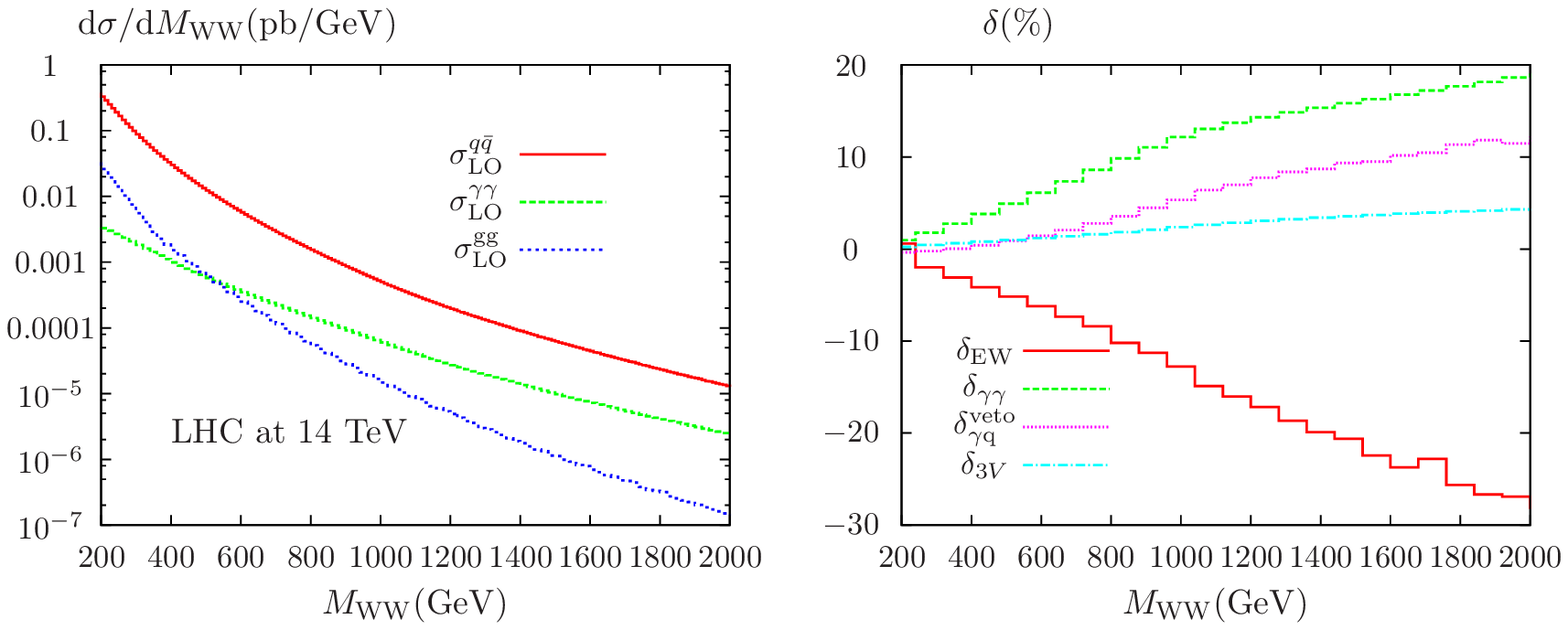}
\includegraphics[width = 0.9\textwidth]{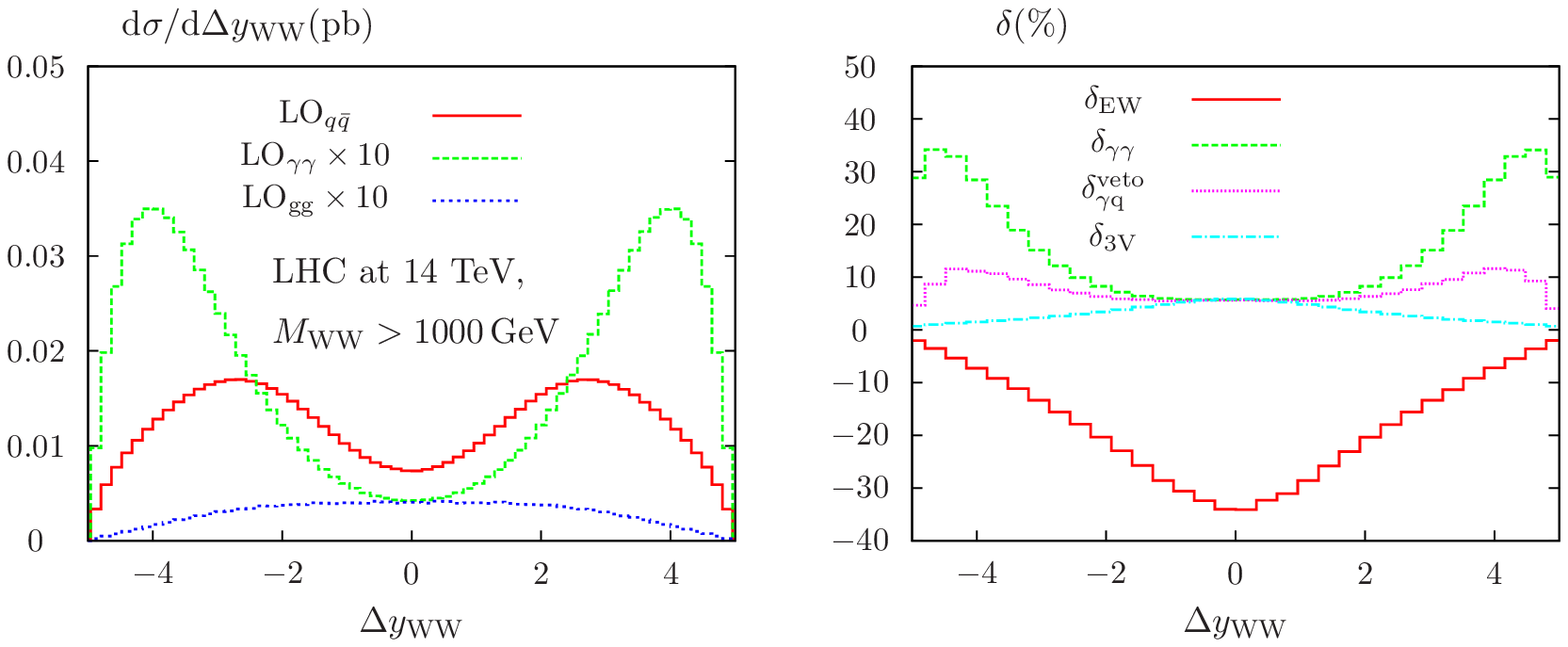}
\caption{Left: Differential LO cross sections to W-pair
  production at LHC14. Right: various EW corrections relative to the
  quark-induced LO process. Top: invariant-mass distribution; Bottom:
  WW rapidity gap for $ M_{\mathrm{WW}} > 1$ TeV. The results presented here
  are obtained in the default setup of
  Ref.~\cite{Bierweiler:2012kw}.}
\label{fi:WWonshell}
\end{center}
\end{figure}
Let us briefly recapitulate the combination of different electroweak
effects in on-shell $V$-pair production~\cite{Bierweiler:2012kw,
  Bierweiler:2013dja,Baglio:2013toa}. In W-pair production, the
invariant-mass distribution (Fig.~\ref{fi:WWonshell} (top)) receives
logarithmically enhanced negative EW corrections
($\delta_{\mathrm{EW}}$). Positive contributions arise from the partonic
subprocess $\gamma\gamma \to \mathrm{WW}$ ($\delta_{\gamma\gamma}$) and
the photon-quark induced processes ($\delta_{\gamma
  q}^{\mathrm{veto}}$), which have been evaluated applying the jet veto
defined in Ref.~\cite{Bierweiler:2012kw}. The effect of massive-boson
radiation ($\delta_{3V}$) is moderate, however, it strongly depends on
the event selection~\cite{Bierweiler:2013dja}.

The above picture significantly changes if angular distributions of the
W-pair are studied at high invariant masses. This can be seen in
Fig.~\ref{fi:WWonshell} (bottom) where distributions of the rapidity gap
of the two Ws are shown for $M_{\mathrm{WW}}>1000$ GeV. While the
genuine EW corrections drastically reduce the differential cross
sections at high $p_{\rT,\mathrm{W}}$, corresponding to small rapidity
gap, the photon-induced contributions significantly increase the rates
at small scattering angles, corresponding to large rapidity gap. As a
result, a dramatic distortion of angular distributions is visible.

The photon-induced corrections presented above suffer
a large systematic error stemming from our ignorance of the photon
content of the proton. This becomes obvious from Fig.~25 of
Ref.~\cite{Ball:2013hta}, where the NNPDF2.3QED~\cite{Ball:2013hta} set
has been used to estimate the error on the $\gamma\gamma$-induced W-pair
cross section. A relative error of $\pm 50\%$ on the
leading-order (LO) cross section at $M_{\mathrm{WW}} = 1000$ GeV is solely
induced by the photon PDF error. This indicates that a significant
improvement in the determination of the photon PDFs is mandatory to
reliably predict the W-pair production cross section at high energies.
Turning to WZ production, the situation is qualitatively similar to WW
production, though here the $\gamma\gamma$ process is absent and the
genuine EW corrections are smaller.  In Z-pair production, however, the
$\gamma q$-induced contributions are negligible, and particularly large
negative corrections dramatically affect Z-pair production at high
transverse momenta.

\subsection{Polarization and decays}
\label{se:off-shell}
\begin{figure}
\begin{center}
\includegraphics[width = 0.9\textwidth]{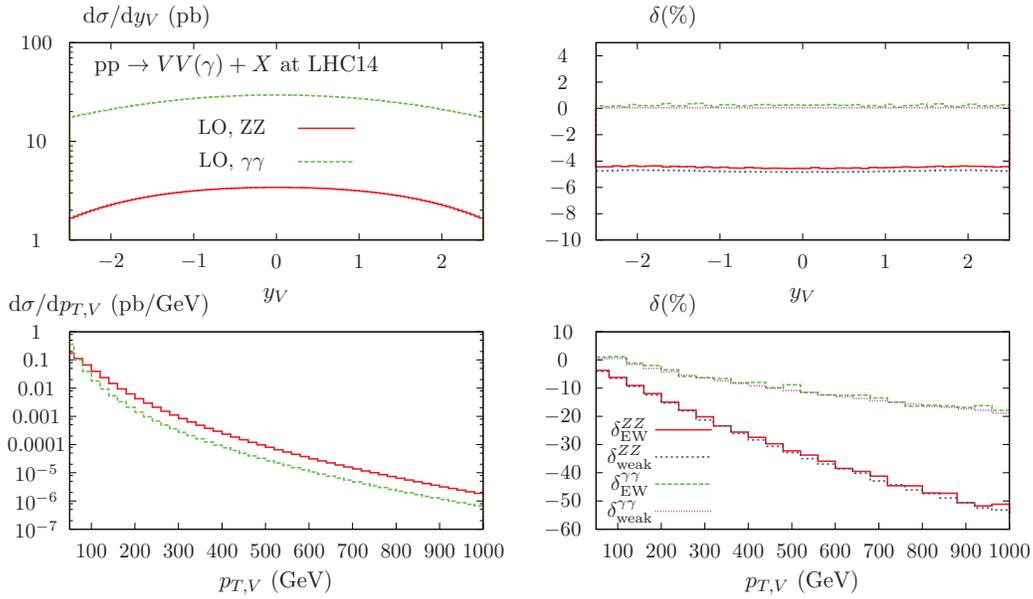}
\caption{Left: Differential LO cross sections to Z- and
  $\gamma$-pair production at LHC14. Right: Full EW corrections and
  purely weak corrections defined relative to the quark-induced LO
  process. Top: vector-boson rapidity; Bottom: vector-boson transverse
  momentum. The results presented here are obtained in the default setup
  of Ref.~\cite{Bierweiler:2013dja}.}\label{fi:ZZonshell}
\end{center}
\end{figure}
Realistic predictions for massive vector-boson pair production require
the inclusion of leptonic decays in the event simulation. Accordingly,
corresponding radiative corrections must be computed for polarized cross
sections to properly include spin correlations.

The purely weak corrections for ZZ (and also $\gamma\gamma$) production
are well defined and the QED contributions can be separated in a
gauge-invariant and infrared-safe way~\cite{Bierweiler:2013dja}, and the
numerical effect of QED corrections in general is below 1\% in all
regions of phase space (see Fig.~\ref{fi:ZZonshell}).  Consequently, it
is well motivated to neglect the QED part in the computation of EW
corrections to Z-pair production.
\begin{figure}
\begin{center}
\includegraphics[width = 0.9\textwidth]{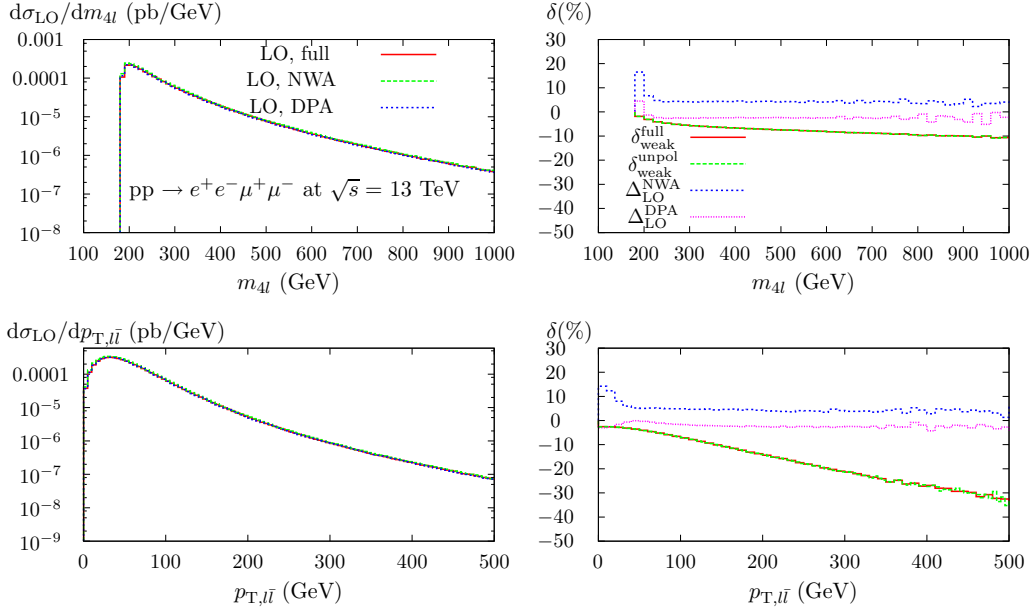}
\caption{ Left: Selected differential LO cross sections for
  process~\protect\eqref{eq:ZZ} at LHC13. Right: relative weak
  corrections with ($\delta_{\mathrm{weak}}^{\mathrm{full}}$) and
  without ($\delta_{\mathrm{weak}}^{\mathrm{unpol}}$) spin correlations
    and relative deviations of the NWA and DPA results from the full LO,
    respectively. Top: four-lepton invariant mass; bottom: Z-boson
    transverse momentum. The results presented have been obtained with
    the CT10 PDF set~\cite{Lai:2010vv}.} \label{fi:ZZoffshell}
\end{center}
\end{figure}
\begin{figure}
\begin{center}
\includegraphics[width = 0.9\textwidth]{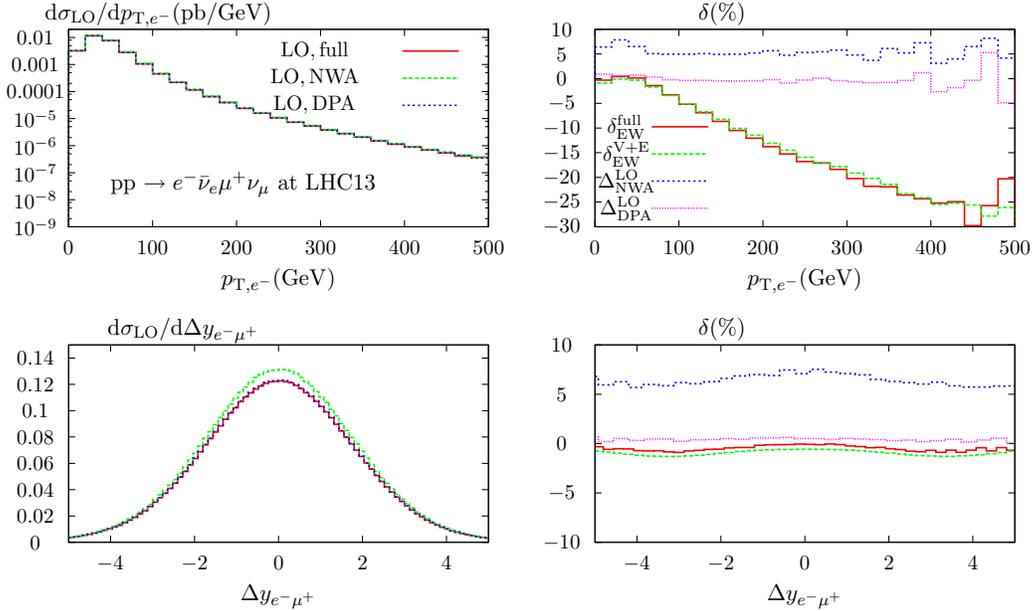}
\end{center}
\caption{\label{fi:WWoffshell} Left: Selected differential LO cross
  sections for process~(\protect\ref{eq:VV}, left) at LHC13. Right: Full
  EW corrections, EW corrections in the V+E approximation and relative
  deviations of the LO NWA and LO DPA results from the full LO,
  respectively. Top: $e^-$ transverse momentum; bottom: charged-lepton
  rapidity gap. The results presented have been obtained with the CT10
  PDF set~\cite{Lai:2010vv}.}
\end{figure}

 \begin{figure}
 \begin{center}
 \includegraphics[width = 0.9\textwidth]{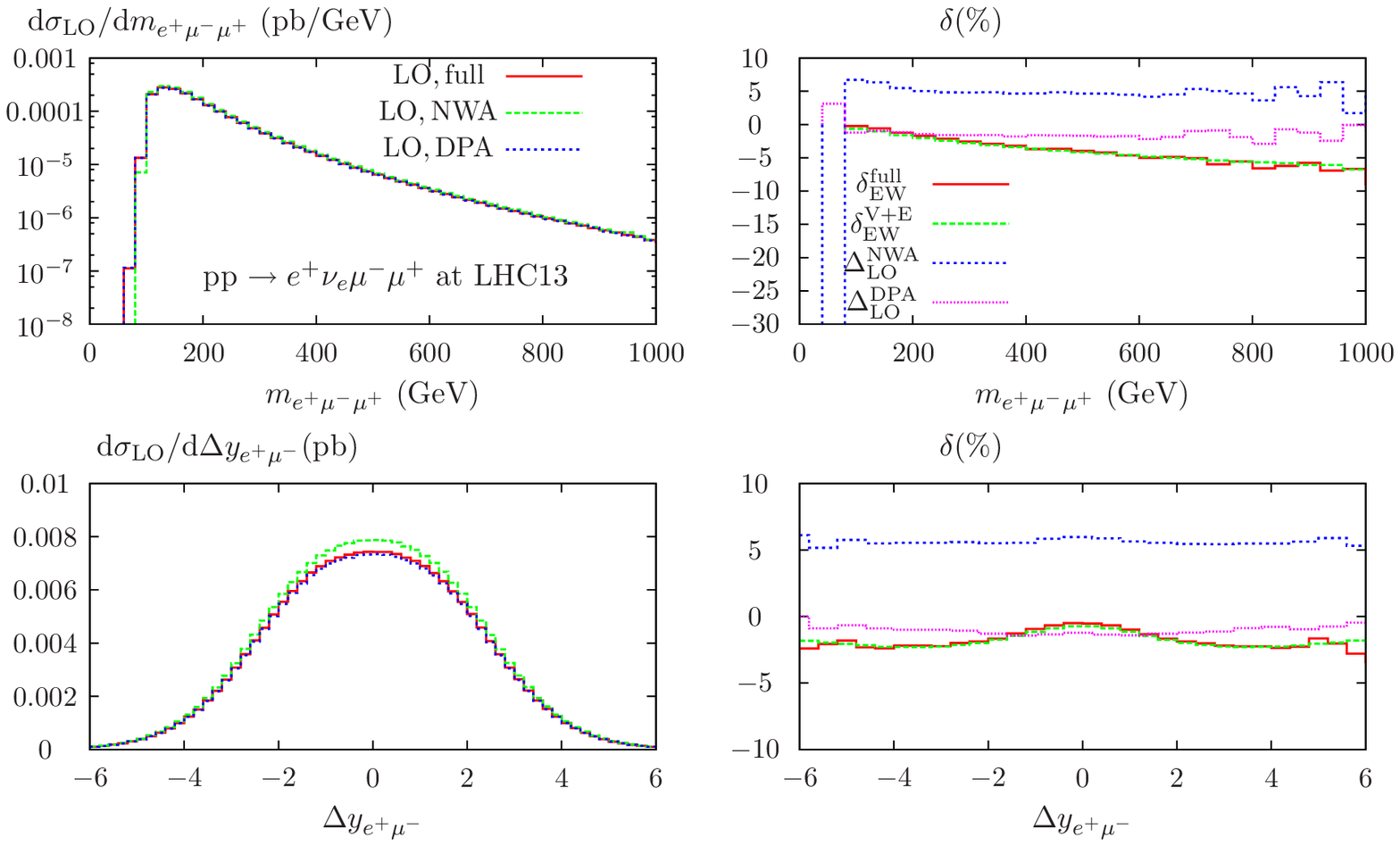}
 \caption{Left: Selected differential LO cross sections for
   process~(\protect\ref{eq:VV}, right) at LHC13. Right: Full EW corrections, EW
   corrections in the V+E approximation and relative deviations of the LO
   NWA and LO DPA results from the full LO, respectively. Top:
   charged-lepton invariant mass; bottom: charged-lepton rapidity
   gap. The results presented have been obtained with the CT10 PDF
   set~\cite{Lai:2010vv}.}
 \label{fi:WZoffshell}
 \end{center}
 \end{figure}

In Fig.~\ref{fi:ZZoffshell} selected LO distributions and corresponding
weak corrections for the process
\begin{equation}
\label{eq:ZZ}
\mathrm{pp} \to (\PZ/\gamma^*)(\PZ/\gamma^*) \to e^+e^-\mu^+\mu^-
\end{equation}
are presented, where $p_{\rT,l}>10\;\mathrm{GeV}$ and $|y_l|<5$ is
required for the charged leptons.  Additionally, the two-lepton
invariant masses are constrained to $|M_{l\bar{l}} - M_{V}| < 25\;
\mathrm{GeV}$ to reduce the admixture of non-resonant contributions and
improve the validity of the pole approximation applied below. The full
LO results obtained in the complex-mass scheme~\cite{Denner:2005fg}
(including also non-resonant and singly-resonant contributions) are well
approximated by the DPA, while in the
narrow-width approximation (NWA) a 10\% discrepancy from the full result
is observed. The relative weak corrections are consistently computed in
the NWA, assuming full factorization of the corrections to production
and decay, respectively. The full spin correlations are accounted for in
the NLO result $\delta_{\mathrm{weak}}^{\mathrm{full}}$.  Final-state
photon radiation from the leptons, which potentially leads to drastic
distortions of leptonic observables, is not included in these results
and will be taken care of in the MC implementation to be discussed in
the next section.

In addition to the full EW corrections we also present corrections
obtained in a simple $K$-factor approach
($\delta_{\mathrm{weak}}^{\mathrm{unpol}}$), where the unpolarized
relative weak corrections $K^{\mathrm{ZZ}}_{q\bar{q}}(\hat{s},\hat{t})$
to the partonic subprocesses $q {\bar q} \to \mathrm{ZZ}$ $(q =
\mbox{u,d,s,c,b})$ have simply been multiplied with the tree-level
result. We observe nearly perfect agreement with the full result,
indicating that spin correlations are hardly affected by weak
corrections~\cite{Bierweiler:2013dja}.

In contrast to the case of Z-pair production, the situation is more
involved in WW or WZ production, where it is not possible to simply
separate the (presumably small) QED corrections. However, the infrared
singularities in the virtual corrections related to photon exchange can
be eliminated by the endpoint contribution as defined in the dipole
subtraction procedure~\cite{Dittmaier:1999mb}. This virtual+endpoint
(V+E) approximation gives a surprisingly good approximation at the level
of a few percent, as clearly visible in Figs.~\ref{fi:WWoffshell}
and~\ref{fi:WZoffshell}, where selected differential distributions for
the processes
\begin{equation}\label{eq:VV}
\mathrm{pp} \to \PW^-\PW^+ \to e^- \bar{\nu}_e\mu^+ \nu_\mu\;\quad
\mbox{and}\quad\mathrm{pp} \to \PW^+(\PZ/\gamma^*) \to e^+ \nu_e \mu^- \mu^+ 
\end{equation}
are presented obtained in the setup defined above, additionally
requiring a minimal missing transverse momentum of 25 GeV. Here,
unpolarized partonic $K$-factors have been used to compute the
corrections $\delta_{\mathrm{EW}}^{\mathrm{V+E}}$ in the NWA using the
V+E approximation; again, good agreement to the full result
$\delta_{\mathrm{EW}}^{\mathrm{full}}$ is observed. As in the ZZ case
one observes that spin correlations are hardly modified by NLO
corrections.

\section{Implementation in HERWIG++}
\label{se:herwig}
\begin{figure}
\begin{center}
\includegraphics[width = 0.45\textwidth]{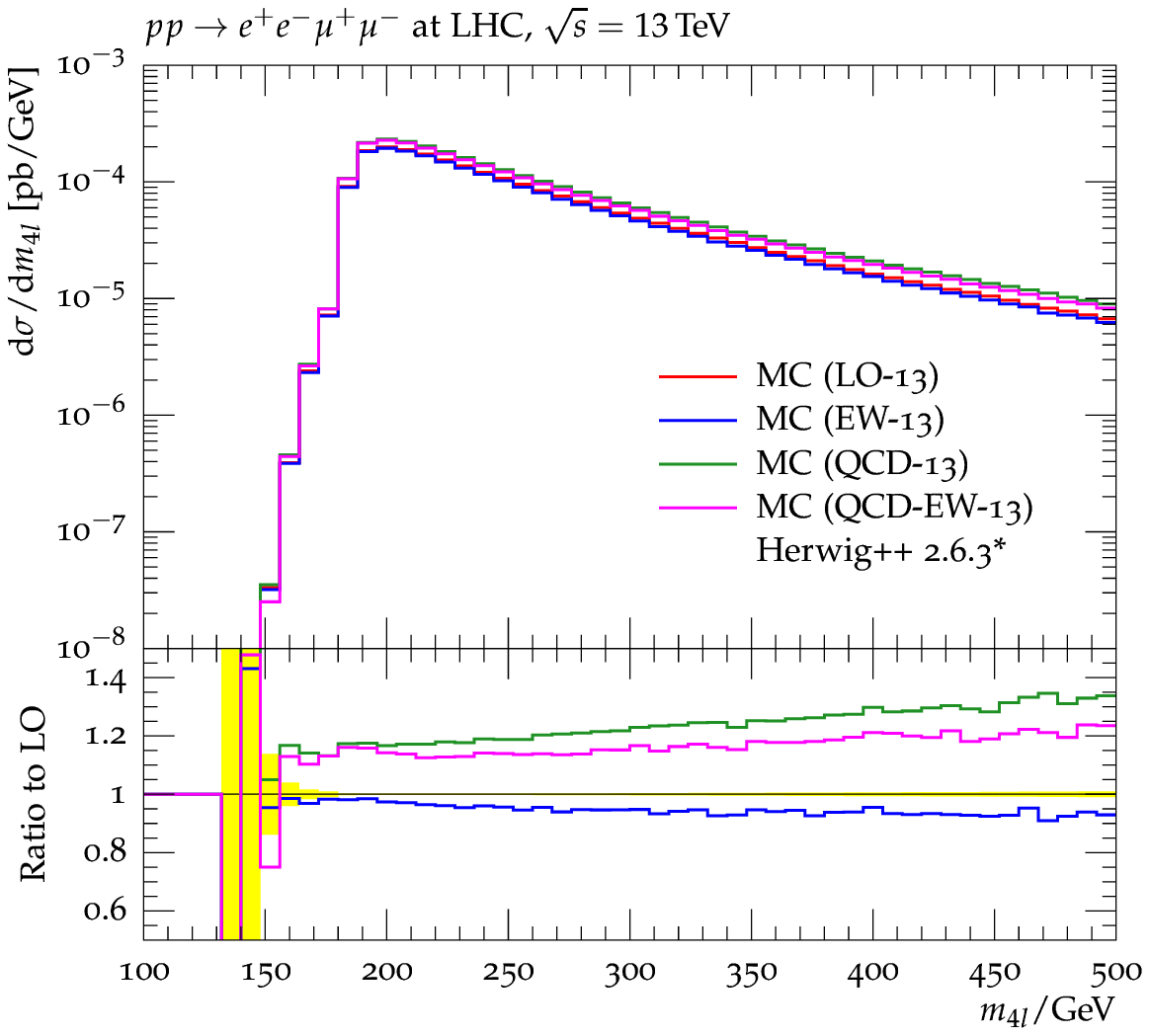}
\hspace{0.5cm}
\includegraphics[width = 0.45\textwidth]{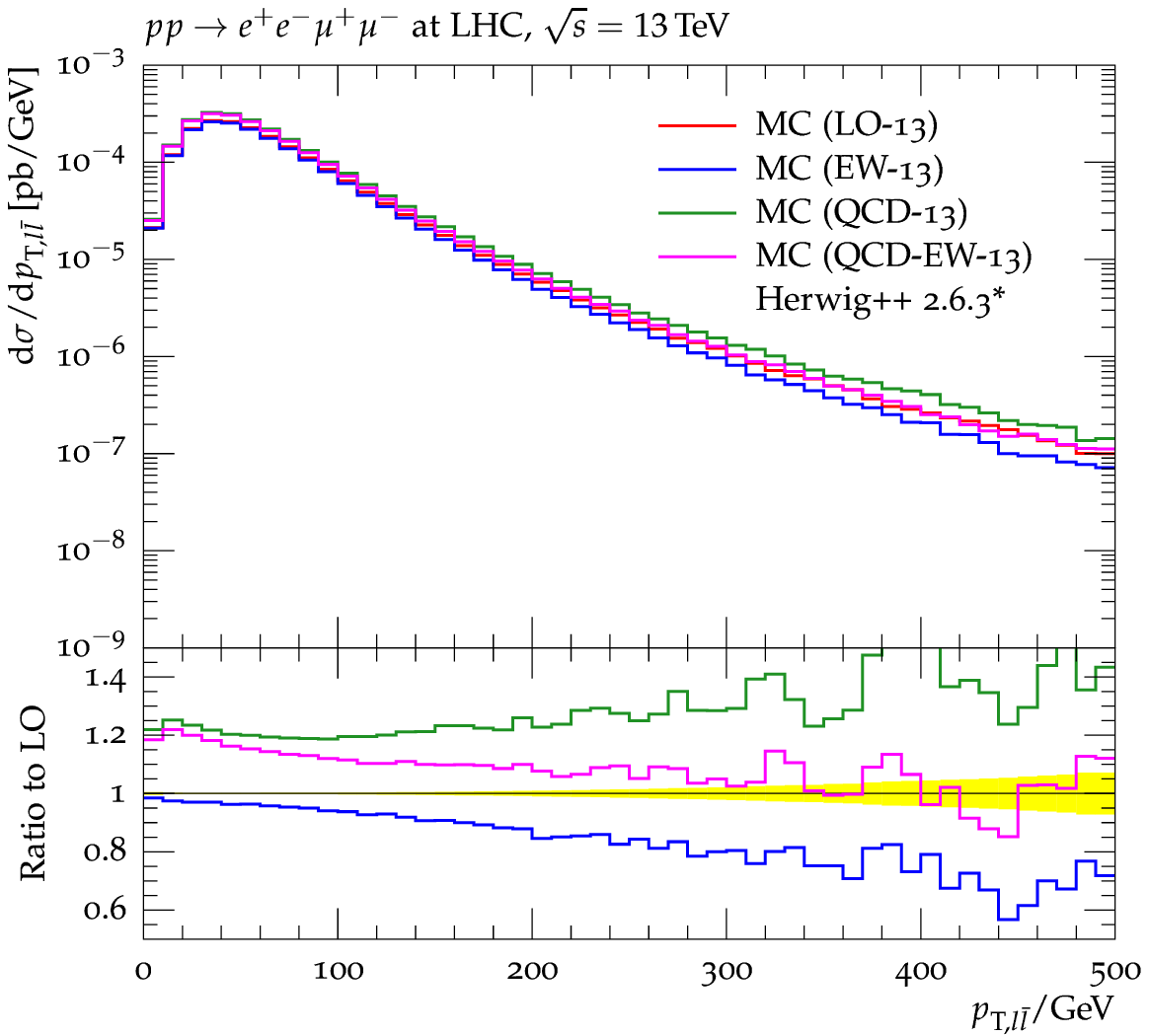}\\
\vspace{0.5cm}
\includegraphics[width = 0.45\textwidth]{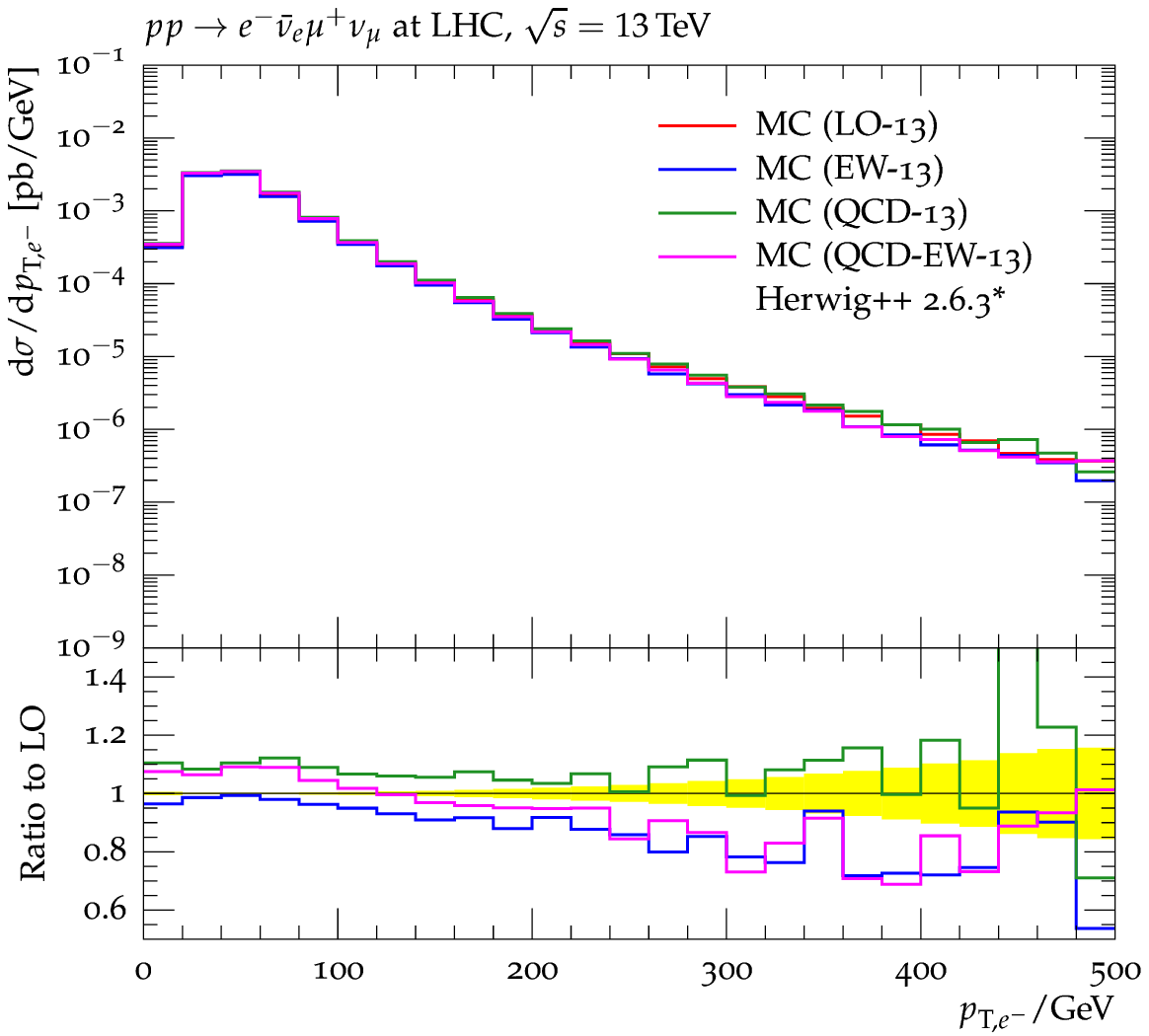}
\hspace{0.5cm}
\includegraphics[width = 0.45\textwidth]{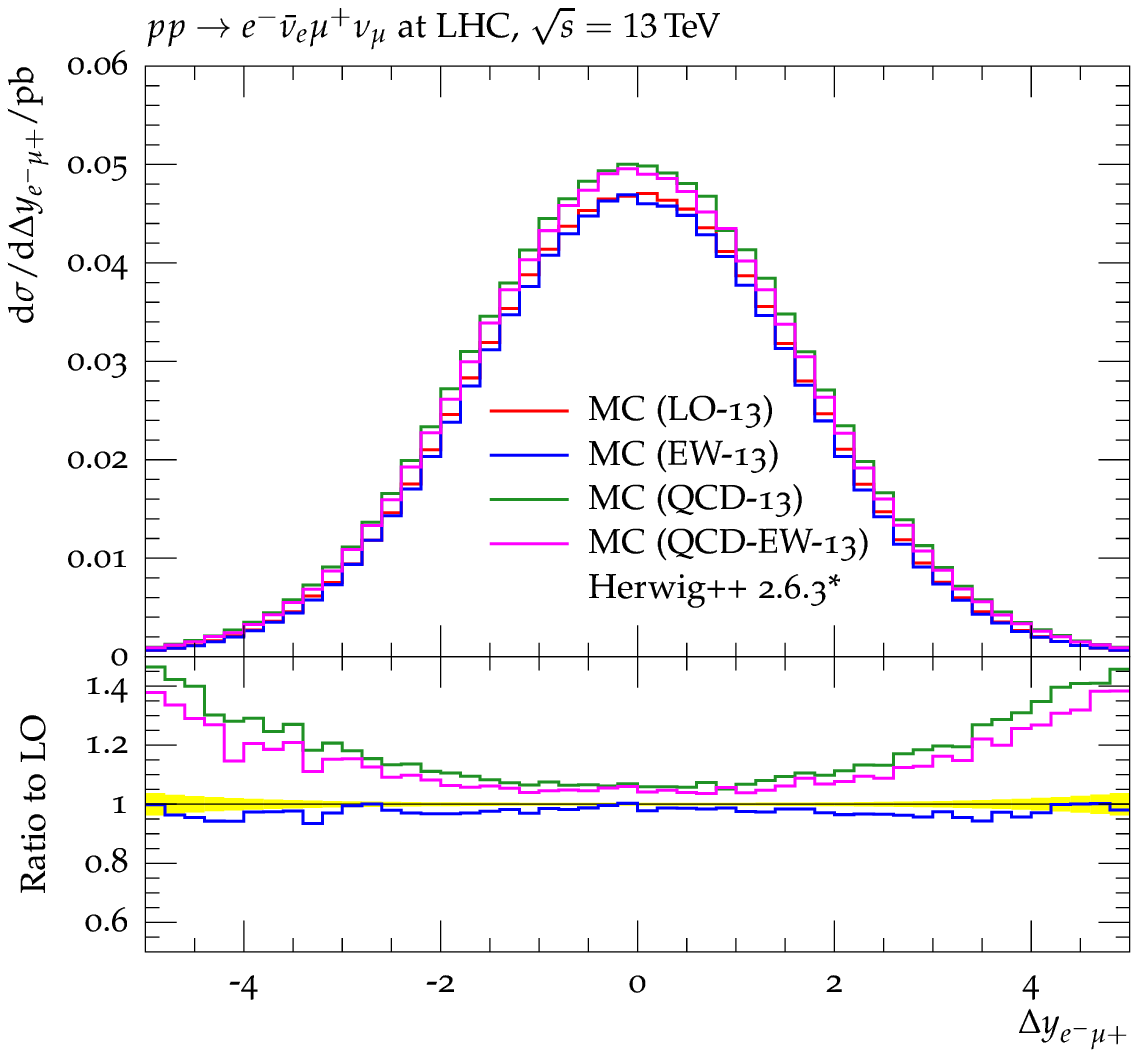}\\
 \vspace{0.5cm}
 \includegraphics[width = 0.45\textwidth]{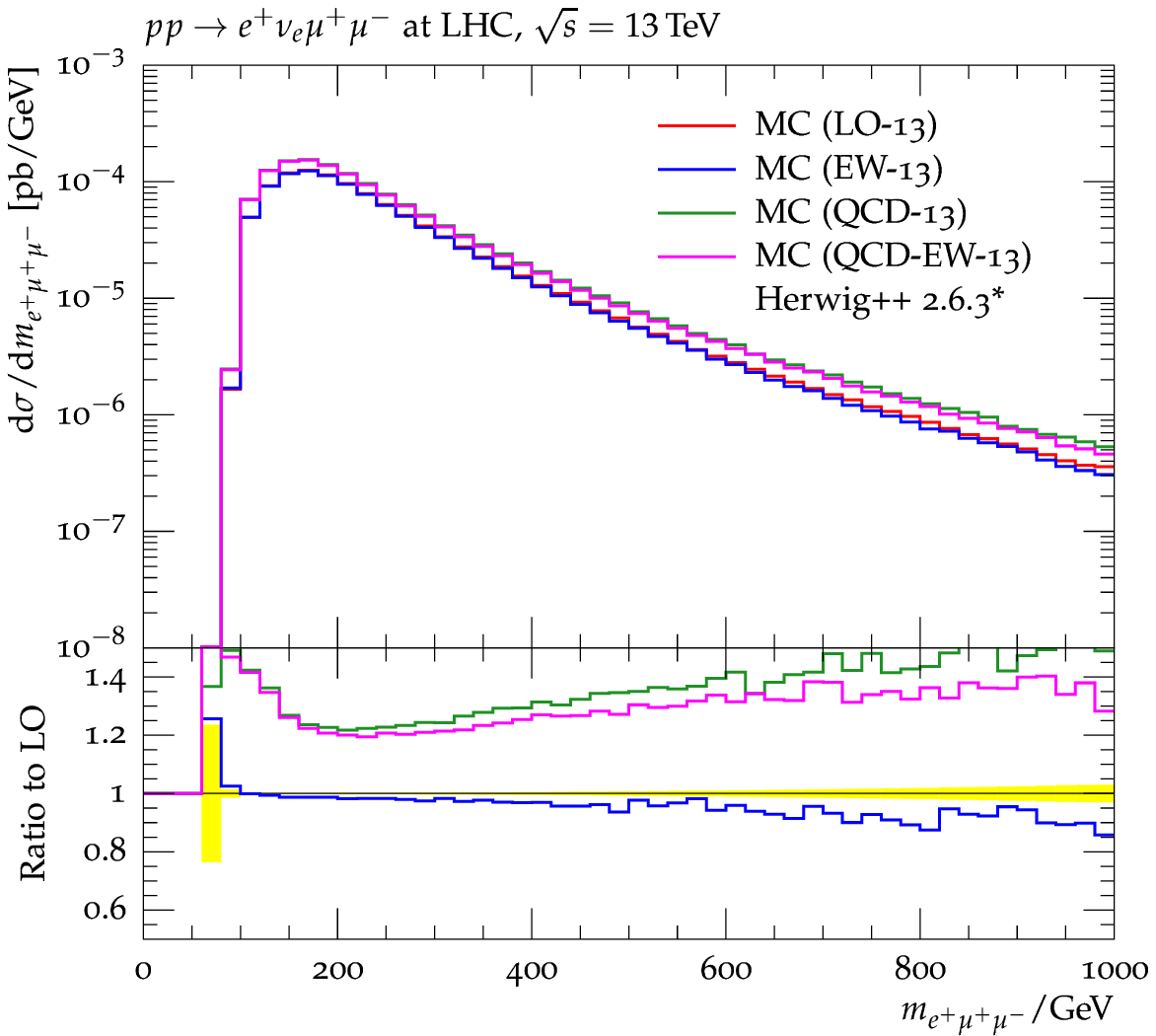}
 \hspace{0.5cm}
 \includegraphics[width = 0.45\textwidth]{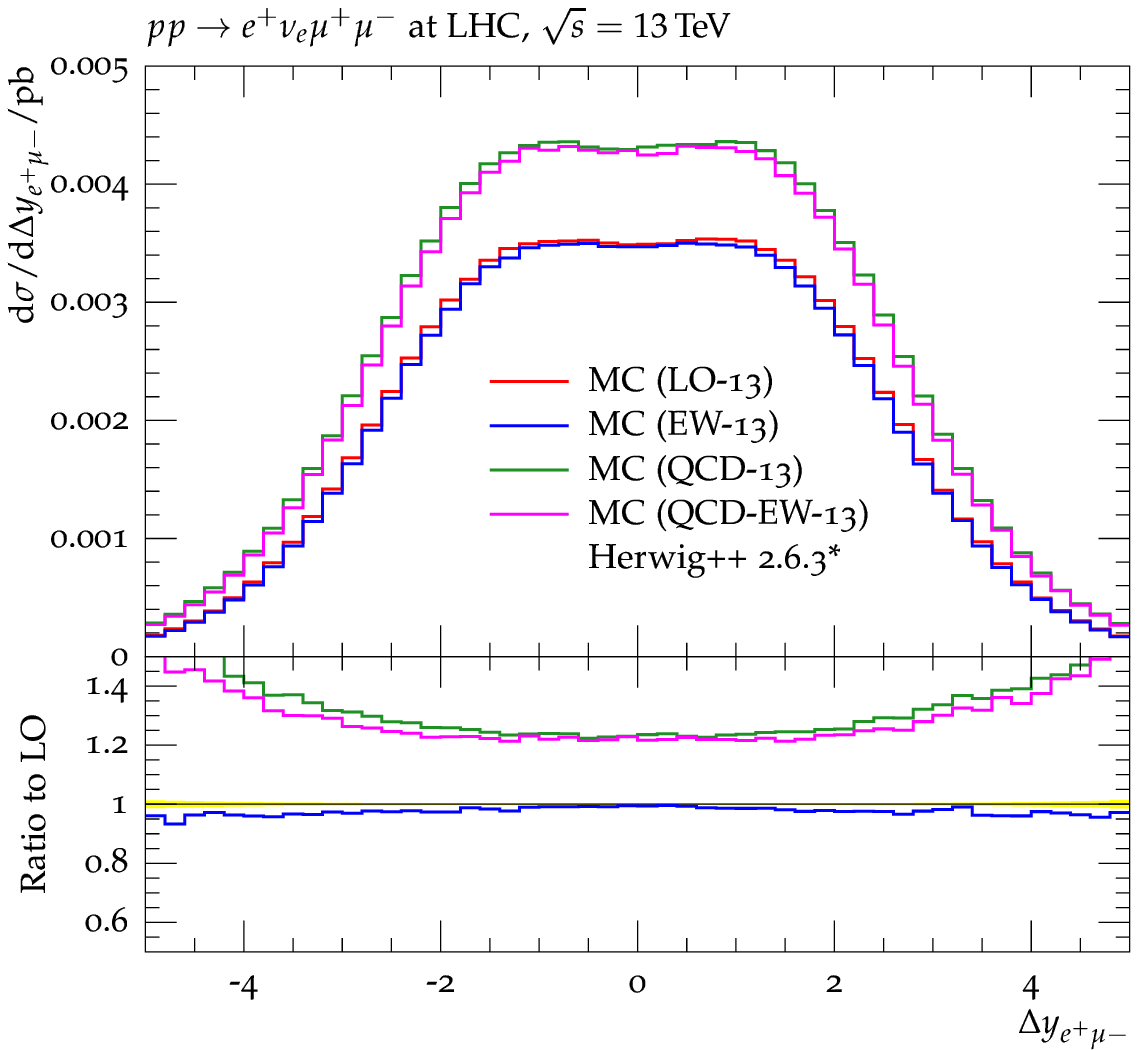}
 \caption{Numerical results for processes~\protect\eqref{eq:ZZ},
   (\protect\ref{eq:VV}, left) and (\protect\ref{eq:VV}, right) obtained
   with the HERWIG++ generator (v2.6.3) for resonant ZZ (top), WW
   (center) and W$^+$Z (bottom) production at LHC13. The results
   directly correspond to those presented in
   Figs.~\protect\ref{fi:ZZoffshell}, \protect\ref{fi:WWoffshell}
   and~\protect\ref{fi:WZoffshell}. In addition, the requirement \mbox{$\left|
     \sum_i \mathbf{p}_{\rT,l}^i \right| <\; 0.3\; \sum_i
   |\mathbf{p}_{\rT,l}^i|$} for balanced vector-boson pairs is included
   in the HERWIG++ predictions.}
\label{fi:herwig}
\end{center}
\end{figure}
Following the philosophy described in the previous section, our MC
implementation of EW corrections is based on partonic unpolarized
$K$-factors $K(\hat{s},\hat{t})$ obtained in the V+E approximation. The
resulting EW correction factor is combined with the respective partonic
cross section provided by the generator in an event-by-event
fashion. The $\hat{s}$ and $\hat{t}$ values are either provided by the
HERWIG++ generator or computed from the final-state lepton momenta.

Since the MC sample generated by HERWIG++ provides, in addition to pure
LO predictions, NLO QCD corrections matched to parton showers as well as
hadronization effects, it is not obvious how to combine the MC cross
sections with partonic EW $K$-factors which are defined relative to a
naive LO. First, it is not clear whether to simply add the EW
corrections on top of the QCD corrected cross section, or to use a
multiplicative ansatz, assuming a factorization of EW and QCD
effects. Second, in the presence of parton showers or NLO QCD
corrections, the simple two-body phase space of the hard $q{\bar q}'\to
V_1V_2$ process, defined by $\hat{s}$ and $\hat{t}$, is distorted by
additional parton radiation, and a proper mapping has to be defined to
restore the two-by-two kinematics.

In our approach, we assume  complete factorization of QCD and EW
corrections. For each event the contribution to the differential partonic 
cross sections is defined by
\begin{equation}
  \rd \hat{\sigma}_{\mathrm{EW} \times \mathrm{QCD}} = K_{q\bar{q}'}^{V_1V_2}(\hat{s},\hat{t}) \times \rd \hat{\sigma}_{\mathrm{QCD}}\,,
\end{equation}
where $\rd \hat{\sigma}_{\mathrm{QCD}}$ denotes the QCD prediction for
$V$-pair production provided by the default HERWIG++
setup~\cite{Hamilton:2010mb}.  The values for $\hat{s}$ and $\hat{t}$
are either directly provided by HERWIG++ or may be computed from the
final-state lepton momenta according to the following prescription: The
squared center-of-mass (CM) energy $\hat{s}'$ is calculated from the
four-lepton final state via $ {\hat s}' = M_{4l}^2\,.  $ The momenta are
boosted into the four-lepton CM frame (denoted by $\Sigma^*$). In this
frame the directions of initial-state hadrons shall be denoted by $
\mathbf{e}^*_i = \frac{\mathbf{p}^*_i}{|\mathbf{p}^*_i|}\,,\; i=1,2\,.
$ The direction of the effective scattering axis in $\Sigma^*$ is now
defined by $ \hat{\mathbf{e}}^* = \frac{\mathbf{e}^*_1 -
  \mathbf{e}^*_2}{|\mathbf{e}^*_1-\mathbf{e}^*_2|}\,, $ and the
effective scattering angle is, correspondingly, given by $ \cos
{\theta}^* = \mathbf{v}^*_1\cdot \hat{\mathbf{e}}^*\,, $ where the
$\mathbf{v}^*_1$ denotes the momentum direction of vector boson
$V_1$. The Mandelstam variable $\hat{t}'$ is then computed from
$\theta^*$ assuming on-shell kinematics. We point out that this
prescription allows for an \emph{a posteriori} implementation of EW
corrections in any MC event sample, provided that the information on the
lepton four-momenta and the IS quark generations are accessible.

Note that the above prescription, as well as the assumption of
factorization of QCD and EW corrections, are only expected to work
reliably if the LO signature of ''balanced'' (i.e.\ back-to-back) vector
bosons is not distorted dramatically by additional high-$p_{\rT}$ QCD
radiation. Therefore, we propose the restriction $\left| \sum_i
  \mathbf{p}_{\rT,l}^i \right| <\; 0.3\; \sum_i |\mathbf{p}_{\rT,l}^i|$
on the $p_{\rT}$ of the visible charged leptons $i$ in the final states
to enforce back-to-back vector bosons and, at the same time, avoid giant
QCD $K$-factors typically occurring at high
$p_\rT$~\cite{Bierweiler:2012kw,Campanario:2012fk}.

Exploiting the above ideas, a simple add-on for EW corrections in the
HERWIG++ setup has been constructed. Numerical results (directly
corresponding to Figs.~\ref{fi:ZZoffshell}, \ref{fi:WWoffshell}
and~\ref{fi:WZoffshell}) are shown in Fig.~\ref{fi:herwig} for resonant
ZZ, WW and W$^+$Z production, respectively. The results presented are
based on samples of 10M events obtained by the HERWIG++ generator
(v2.6.3)~\cite{Arnold:2012fq}. The relative corrections (QCD, EW,
EW$\times$QCD) are normalized to the LO prediction, which includes
parton showers and hadronization effects. In addition, all predictions
also include final-state photon radiation which is implemented by
default in the HERWIG++ setup~\cite{Hamilton:2006xz}. One observes good
agreement between the HERWIG++ ratios and the relative corrections from
Figs.~\ref{fi:ZZoffshell}, \ref{fi:WWoffshell} and~\ref{fi:WZoffshell},
giving a proof of principle that our implementation works reliably.

\section{Conclusions}
\label{se:concl}
An analysis of EW corrections to vector-boson pair production at the LHC
is presented, including the full mass dependence as well as leptonic
decays. Hence, our predictions are valid in the whole kinematic regime
accessible at the LHC. Combining EW corrections and photon-induced
contributions, drastic distortions of angular distributions at high
energies are observed which could easily be misinterpreted as signals of
anomalous couplings. For this reason, and in view of the future
high-luminosity run of the LHC at a CM energy of \mbox{13 TeV}, the
inclusion of EW corrections in the analysis of experimental data by
means of MC techniques is indispensable. Thus we have proposed a simple
method to combine EW corrections with any multi-purpose MC generator. To
demonstrate the practicability of the method, phenomenological results
are presented on the basis of a HERWIG++ event sample, combining EW
corrections with state-of-the-art QCD predictions.

\subsection*{Acknowledgements}
\noindent
Supported by SFB TR9 ``Computational and Particle
Physics'' and BMBF Contract 05HT4VKATI3.

\end{document}